\documentclass{article}
\usepackage{amssymb}

\input{tcilatex}
\begin{document}

\begin{center}
\textbf{Brans-Dicke wormholes in the Jordan and Einstein frames}

K. K. Nandi$^{1}$, B. Bhattacharjee$^{2}$, S. M. K. Alam$^{1,3}$ and J. Evans%
$^{4}$

$^{1}$Department of Mathematics, University of North Bengal, Darjeeling
(W.B.), 734 430, India

$^{2}$Department of Physics, University of North Bengal, Darjeeling (W.B.),
734 430, India

$^{3}$265/3, West Sheorapara, Mirpur, Dhaka, Bangladesh

$^{4}$Department of Physics, University of Puget Sound, Tacoma, Washington
98416

\textbf{Abstract}
\end{center}

We examine the possibility of static wormhole solutions in the vacuum
Brans-Dicke theory both in the original (Jordan) frame and in the
conformally rescaled (Einstein) frame. It turns out that, in the former
frame, wormholes exist only in a very narrow interval of the coupling
parameter, viz., $-3/2<\omega <-4/3$. It is shown that these wormholes are
not traversable in practice. In the latter frame, wormhole solutions do not
exist at all unless energy conditions are violated by hand.


\begin{center}
\textbf{I. INTRODUCTION}
\end{center}

Over the last few years, considerable interest has grown in the field of
wormhole physics, following especially the seminal works of Morris, Thorne,
and Yurtsever [1,2]. Wormholes are topology changes that connect two
asymptotically flat regions. Potential applications of wormhole physics
range from the interpretation of gravitational lensing effects to the
resolution of several outstanding problems in cosmology [3--5].

In the context of traversable wormholes, a crucial issue is the constraint
upon the violation of energy conditions by the stress tensor of quantum or
classical fields. There exist several pointwise and average energy
conditions [6]. Specifically, for quantum fields, Ford and Roman [7] have
proposed, on the basis of certain assumptions, an inequality that constrains
the magnitude of the negative energy density at the throat of a traversable
wormhole. A fundamental assumption for quantum wormholes is that the stress
energy of the spacetime is a renormalized expectation value of the
energy-momentum operator in some quantum state, say, $\left\vert \psi
\right\rangle $. In the literature [8], one actually considers field
equations of semiclassical gravity in the form $G_{\mu \nu }=8\pi <\psi
\left\vert T_{\mu \nu }\right\vert \psi >$. However, some doubts have been
raised, notably by Unruh [9], as to whether field equations in this form
could be an exact description of gravity [10]. On the other hand, quantized
source fields obey well-defined uncertainty relations and it is expected
that uncertainty in the source would induce uncertainty in the gravidynamic
variables and in the light cone structure of spacetime [11,12]. If the
source is taken as $\left\langle T_{\mu \nu }\right\rangle $, such
fluctuations would not occur. Despite these questions, it must be emphasized
that field equations in the above form provide a very good approximation in
many physical situations, especially in the description of the early
universe [13].

There also exist classical fields playing the role of \textquotedblleft
exotic matter\textquotedblright\ that violates the weak energy condition
(WEC), at least at the throat of the wormhole. Examples are provided by the
stress-energy tensors occurring in theories where the action contains $%
\mathbf{R}+\mathbf{R}^{2}$ terms [14], an antisymmetric 3-form axion field
coupled to scalar fields [15], and minimally coupled fields with a
self-interacting potential [16]. Other theories include string-inspired
four-dimensional gravity coupled nonminimally to a scalar field [17], Zee's
induced gravity [18], and the Brans-Dicke scalar-tensor theory [19]. Most of
the works concentrate on dynamic wormholes, while work on static wormholes
is relatively scarce. In particular, in the Brans-Dicke theory, a search for
static wormholes has been initiated only recently [20,21], followed by
Anchordoqui, Bergliaffa, and Torres [22]. Considering the importance of
Brans-Dicke theory in the interpretation of various physical phenomena
[23--25] and owing to the fact that, in the limit $\omega \rightarrow \infty
$, one recovers general relativity, it is only desirable that a thorough
study of classical wormhole solutions be undertaken in this theory.

In this paper, we intend to examine wormhole solutions in the Jordan and
Einstein frames which are defined as follows [26]: The pair of variables
(metric $g_{\mu \nu }$, scalar $\varphi $) defined originally in the
Brans-Dicke theory constitute what is called a Jordan frame. Consider now a
conformal rescaling%
\begin{equation}
\widetilde{g}_{\mu \nu }=f(\varphi )g_{\mu \nu }\text{, \ \ \ \ \ \ \ }\phi
=g(\varphi )
\end{equation}%
such that, in the redefined action, $\phi $ becomes minimally coupled to $%
\widetilde{g}_{\mu \nu }$ for some functions $f(\varphi )$ and $g(\varphi )$%
. Then the new pair ($\widetilde{g}_{\mu \nu },\phi $) is said to constitute
an Einstein frame. There exist different viewpoints as to the question of
which of these two frames is physical, but the arguments of Magnano and Soko%
\l owski [26] seem convincing enough in favor of the physicality of the
Einstein frame.

In what follows, we shall be concerned only with static spherically
symmetric solutions of the Brans-Dicke theory. For this purpose, only a
class I type of solution is considered; other classes (II--IV) of solutions
can be dealt with in a similar way. Our results are stated as follows. In
Sec. II, we consider the Jordan frame and derive the general condition for
the existence of wormholes. This condition is then used to find wormhole
ranges of $\omega $ in specific cases. Section III shows that these
wormholes are not traversable due to the occurrence of a naked singularity.
The Einstein frame is considered in Sec. IV, and it is shown that wormhole
solutions do not exist at all in that frame. The last section, Sec. V, is a
summary.

\begin{center}
\textbf{II. JORDAN FRAME}
\end{center}

In order to investigate the possibility of wormholes in the vacuum
(matter-free) Brans-Dicke theory, it is convenient to cast the spacetime
metric in the Morris-Thorne canonical form%
\begin{equation}
d\tau ^{2}=-e^{2\Phi (R)}dt^{2}+\left[ 1-\frac{b(R)}{R}\right]
^{-1}dR^{2}+R^{2}(d\theta ^{2}+\sin ^{2}\theta d\psi ^{2})
\end{equation}%
where $\Phi (R)$ and $b(R)$ are redshift and shape functions, respectively.
These functions are required to satisfy some constraints, enumerated in [1],
in order that they represent a wormhole. It is, however, important to stress
that the choice of coordinates (Morris-Thorne) is purely a matter of
convenience and not a physical necessity. For instance, one could equally
well work directly with isotropic coordinates using the analysis of Visser
[6], but the final conclusions would be the same. Nonetheless, it must be
understood that a more appropriate procedure should involve coordinate
independent proper quantities.

The matter-free action in the Jordan variables is ($G=c=1$)%
\begin{equation}
S=\frac{1}{16\pi }\int d^{4}x(-g)^{\frac{1}{2}}\left[ \varphi \mathbf{R}%
-\varphi ^{-1}\omega (\varphi )g^{\mu \nu }\varphi _{,\mu }\varphi _{,\nu }%
\right] .
\end{equation}%
The field equations are%
\[
\square ^{2}\varphi =0,
\]

\begin{equation}
\mathbf{R}_{\mu \nu }-\frac{1}{2}g_{\mu \nu }\mathbf{R}=-\frac{\omega }{%
\varphi ^{2}}\left[ \varphi _{,\mu }\varphi _{,\nu }-\frac{1}{2}g_{\mu \nu
}\varphi _{,\sigma }\varphi ^{,\sigma }\right] -\frac{1}{\varphi }\left[
\varphi _{;\mu }\varphi _{;\nu }-g_{\mu \nu }\square ^{2}\varphi \right] ,
\end{equation}%
where $\square ^{2}\equiv (\varphi ^{;\rho })_{;\rho }$ and $\omega $ is a
dimensionless coupling parameter. The general solution, in isotropic
coordinates ($t,r,\theta ,\varphi $), is given by%
\begin{equation}
d\tau ^{2}=-e^{2\alpha (r)}dt^{2}+e^{2\beta (r)}dr^{2}+e^{2\nu
(r)}r^{2}(d\theta ^{2}+\sin ^{2}\theta d\psi ^{2}).
\end{equation}%
Brans class I solutions [27] correspond to the gauge $\beta -\nu =0$ and are
given by%
\begin{equation}
e^{\alpha (r)}=e^{\alpha _{0}}\left[ \frac{1-B/r}{1+B/r}\right] ^{\frac{1}{%
\lambda }},
\end{equation}%
\begin{equation}
e^{\beta (r)}=e^{\beta _{0}}\left[ 1+B/r\right] ^{2}\left[ \frac{1-B/r}{1+B/r%
}\right] ^{\frac{\lambda -C-1}{\lambda }},
\end{equation}%
\begin{equation}
\varphi (r)=\varphi _{0}\left[ \frac{1-B/r}{1+B/r}\right] ^{\frac{C}{\lambda
}},
\end{equation}%
\begin{equation}
\lambda ^{2}\equiv (C+1)^{2}-C\left( 1-\frac{\omega C}{2}\right) >0,
\end{equation}%
where $\alpha _{0}$, $\beta _{0}$, $B$, $C$, and $\varphi _{0}$ are
constants. The constants $\alpha _{0}$ and $\beta _{0}$ are determined by
asymptotic flatness condition as $\alpha _{0}=$ $\beta _{0}=0$.

Redefining the radial coordinate $r\rightarrow R$ in the metric (5) as%
\begin{equation}
R=re^{\beta _{0}}\left[ 1+B/r\right] ^{2}\left[ \frac{1-B/r}{1+B/r}\right]
^{\Omega }\text{, \ \ \ \ \ \ \ \ \ }\Omega =1-\frac{C+1}{\lambda },
\end{equation}%
we obtain the following functions for $\Phi (R)$ and $b(R)$:%
\begin{equation}
\Phi (R)=\alpha _{0}+\frac{1}{\lambda }\left[ \ln \left\{ 1-\frac{B}{r(R)}%
\right\} -\ln \left\{ 1+\frac{B}{r(R)}\right\} \right] ,
\end{equation}%
\begin{equation}
b(R)=R\left[ 1-\left\{ \frac{\lambda \{r^{2}(R)+B^{2}\}-2r(R)B(C+1)}{\lambda
\{r^{2}(R)+B^{2}\}}\right\} ^{2}\right] .
\end{equation}%
The throat of the wormhole occurs at $R=R_{0}$ such that $b(R_{0})=R_{0}$.
This gives minimum allowed $r$-coordinate radii $r_{0}^{\pm }$ as%
\begin{equation}
r_{0}^{\pm }=\alpha ^{\pm }B,
\end{equation}%
\begin{equation}
\alpha ^{\pm }=(1-\Omega )\pm \sqrt{\Omega (\Omega -2)}.
\end{equation}%
The values $R_{0}^{\pm }$ can be obtained from Eq. (10) using this $%
r_{0}^{\pm }$. Noting that $R\rightarrow \infty $ as $r\rightarrow \infty $,
we find that $b(R)/R\rightarrow 0$ as $R\rightarrow \infty $. Also, $%
b(R)/R\leq 1$ for all $R\geq $ $R_{0}^{\pm }$. The redshift function $\Phi
(R)$ has a singularity at $r=r_{S}=B$. In order that a wormhole be just
geometrically traversable, the minimum allowed values $r_{0}^{\pm }$ must
exceed $r_{S}=B$. It can be immediately verified from Eq. (10) that $%
r_{0}^{\pm }\geq B\Rightarrow $ $R_{0}^{\pm }\geq 0$. This is possible only
if the range of $\Omega $ is chosen either as $-\infty <\Omega \leq 0$ or as
$2<\Omega <\infty $. We shall not consider the latter range here.

The energy density of the wormhole material is given by [1]%
\begin{equation}
\rho (R)=(8\pi R^{-2})(db/dR)
\end{equation}%
and a straightforward calculation gives%
\[
db/dR=4B^{2}r^{2}(R)[r^{2}(R)-B^{2}]^{-2}\Omega (2-\Omega )
\]%
\begin{equation}
=4B^{2}r^{2}(R)[r^{2}(R)-B^{2}]^{-2}\left[ 1-\left( \frac{C+1}{\lambda }%
\right) ^{2}\right] .
\end{equation}%
Therefore, the most general condition for the violation of the WEC is that

\begin{equation}
C(\omega )+1>\lambda (\omega ),
\end{equation}
where the real function $C(\omega )$ is as yet unspecified. (Strictly
speaking, the condition should read $[C(\omega )+1]^{2}>\lambda ^{2}(\omega
) $, see the discussion in Refs. [40], [41]). However, as long as the
special condition (17), which ensures $R_{0}^{\pm }>0$, is satisfied, it
follows that%
\begin{equation}
b_{0}^{\prime }=\frac{db}{dR}\mid _{R=R_{0}^{\pm }}=-1,
\end{equation}%
so that $\rho _{0}=\rho \mid _{R=R_{0}^{\pm }}<0$, and a violation of the
WEC at the throat is achieved thereby. In the limit $r_{0}^{\pm }$.$%
\rightarrow B+$, or, equivalently, $R_{0}^{\pm }\rightarrow 0+$, one obtains
$\rho _{0}\rightarrow -\infty $. This means that there occurs an infinitely
large concentration of exotic matter at the throat when its $r$ radius is in
the vicinity of the Schwarzschild radius $r_{S}=B$. No upper limit to this
classical negative energy density is known to us. The general profile for $%
\rho (R)$ for a given wormhole configuration is that $\rho (R)$ attains its
maximum at the throat and falls off in an inverse square law as one moves
away from the throat to the asymptotic region.

The constraint (17) can be rephrased, using Eq. (9), as%
\begin{equation}
C(\omega )\left[ 1-\frac{\omega C(\omega )}{2}\right] >0,
\end{equation}%
and depending on the form of $C(\omega )$, this inequality fixes the range
of wormhole values of $\omega $, provided one excludes the forbidden range
coming from the requirement that $\lambda ^{2}>0$. A further exclusion of
the range $\omega \leq -3/2$ comes from a \textquotedblleft
physical\textquotedblright\ requirement that the theory be transferrable to
Einstein frame [26]. In the limiting case, $C(\omega )\rightarrow 0$, $%
\lambda (\omega )\rightarrow 1$ as $\omega \rightarrow \infty $, one simply
recovers the Schwarzschild exterior metric in standard coordinates from Eqs.
(11) and (12), so that $b(R)=2M$ and $b_{0}^{\prime }=0$. The inequality
(19) is violated, and there occurs no traversable wormhole, as is well known
[1].

The analysis of Agnese and La Camera [20] corresponds, as pointed out
earlier [21], to the choice%
\begin{equation}
C(\omega )=-\frac{1}{\omega +2},
\end{equation}%
which suggests, via Eq.(19), a wormhole range $\omega <-4/3$. The forbidden
range turns out to be $-2<\omega <-3/2$, which is already a part of the
unphysical range $\omega \leq -3/2$. Therefore, one is left with a very
narrow actual interval for wormhole solutions, viz., $-3/2<\omega <-4/3$. It
appears that the authors just missed this interval.

We should recall here that Eq. (20) is derived on the basis of a weak field
(post Newtonian) approximation and there is no reason for Eq. (20) to hold
for stars with a strong field such as neutron stars. In reality, if we
assume such a restriction as Eq. (20), the junction conditions for the
metric and scalar field are not satisfied at the boundary of the stars [28].
Evidently, any form for $C(\omega )$ different from Eq. (20) would lead to a
different wormhole interval for v. For example, in the context of
gravitational collapse in the Brans-Dicke theory, Matsuda [28] chose $%
C(\omega )\varpropto -\omega ^{-1/2}$. Let us take $C(\omega )=-q\omega
^{-1/2}$ and choose $q<0$ such that $C(\omega )>0$. Then the constraint (19)
will be satisfied only if $\omega >4/q^{2}$ The exact form of $C(\omega )$
should be known \textit{a priori} from other physical considerations.
However, this is just a tentative example and is meant to highlight how
crucially the wormhole range for $\omega $ depends on the form of $C(\omega
) $.

The constraint (17) is based only on the requirement of geometric
traversability, i.e., on the requirement that the throat radii be larger
than the event horizon radius $r=B$. Therefore, an immediate inquiry is
whether such wormholes are traversable in practice. We discuss this issue in
the following section.

\begin{center}
\textbf{III. TRAVERSABILITY}
\end{center}

In order to get a first hand idea about traversability in the Jordan frame,
a convenient procedure is to calculate the scales over which wormhole
functions change. Ford and Roman [7] defined the following quantities at the
throat $R=R_{0}$ of a traversable wormhole:%
\begin{equation}
\overline{r}_{0}=R_{0},r_{1}=\frac{R_{0}}{\left\vert b_{0}^{\prime
}\right\vert },R_{2}=\frac{1}{\left\vert \Phi _{0}^{\prime }\right\vert }%
,r_{3}=\left\vert \frac{\Phi _{0}^{\prime }}{\Phi _{0}^{\prime \prime }}%
\right\vert .
\end{equation}%
These quantities are a measure of coordinate length scales at the throat
over which the functions $b(R)$, $\Phi (R)$, and $\Phi ^{\prime }(R)$
change, respectively. For the class I solutions, they become%
\begin{equation}
\overline{r}_{0}=R_{0}^{\pm },r_{1}=R_{0}^{\pm },R_{2}=0,r_{3}=0.
\end{equation}%
The vanishing of $R_{2}$ and $r_{3}$ implies that both $\Phi (R)$ and $\Phi
^{\prime }(R)$ exhibit an abrupt jump at the throat. It is therefore
expected that the tidal forces at the throat would be large. That this is
indeed so can be verified by calculating, for example, the differential of
the radial tidal acceleration [1] given in an orthonormal frame ($\widehat{e}%
_{t,}\widehat{e}_{R,}\widehat{e}_{\theta ,}\widehat{e}_{\varphi }$) by%
\begin{equation}
\Delta a^{r}=-\mathbf{R}_{\widehat{R}\widehat{t}\widehat{R}\widehat{t}}\xi
^{R},
\end{equation}%
where $\xi ^{R}$ is the radial component of the separation vector and%
\begin{equation}
\left\vert \mathbf{R}_{\widehat{R}\widehat{t}\widehat{R}\widehat{t}%
}\right\vert =\left\vert (1-b/R)\left\{ -\Phi ^{\prime \prime }+\frac{%
b^{\prime }R-b}{2R(R-b)}-(\Phi ^{\prime })^{2}\right\} \right\vert .
\end{equation}%
For the metric given by Eqs. (11) and (12), we find%
\begin{equation}
\left\vert \mathbf{R}_{\widehat{R}\widehat{t}\widehat{R}\widehat{t}%
}\right\vert =\left\vert \frac{Br}{\lambda R^{2}(r^{2}-B^{2})}\left\{
2(1-b/R)^{1/2}+(1-b/R)^{-1/2}b^{\prime }+\frac{\lambda (r^{2}+B^{2})-4Br}{%
\lambda (r^{2}-B^{2})}\right\} \right\vert .
\end{equation}%
At the throat where $b(R_{0}^{\pm })=R_{0}^{\pm }$, we have $\left\vert
\mathbf{R}_{\widehat{R}\widehat{t}\widehat{R}\widehat{t}}\right\vert
\rightarrow $ a finite value, not $\infty $ (a correction on the printed
paper) and this implies . As we march away from the throat to the asymptotic
limit $r\rightarrow \infty $ or, $R\rightarrow \infty $, we find $\left\vert
\mathbf{R}_{\widehat{R}\widehat{t}\widehat{R}\widehat{t}}\right\vert
\rightarrow 0$, as is to be expected. However, $\left\vert \mathbf{R}_{%
\widehat{R}\widehat{t}\widehat{R}\widehat{t}}\right\vert \rightarrow \infty $
as $r\rightarrow B+$.

Such an infinitely large tidal force at the horizon is presumably related to
the presence of singular null surface or naked singularity in the wormhole
spacetime. These wormholes, to use a phrase by Visser [6], are
\textquotedblleft badly diseased\textquotedblright . The occurrence of
singular null surface in the scalar tensor theories is directly related to
the \textquotedblleft no-hair theorem,\textquotedblright\ which commonly
means that \textquotedblleft black holes have no scalar
hair\textquotedblright\ [29]. Early investigations into the no-hair theorem
in the Brans-Dicke theory are due to Hawking [30], Chase [31], Teitelboim
[32], and Bekenstein [33]. Recently, Saa [34] has formulated a new no-hair
theorem which basically relies on the assessment of the behavior of scalar
curvature R, which, for the metric (6) and (7), turns out to be%
\begin{equation}
\mathbf{R}(r)=\frac{4\omega C^{2}B^{2}r^{4}(r+B)^{2\Omega -6}}{\lambda
^{2}(r-B)^{2\Omega +2}}.
\end{equation}%
Then it follows that $\mathbf{R}\rightarrow \infty $ as $r\rightarrow B+$
for $C\neq 0$. In other words, the scalar curvature diverges as $%
R\rightarrow 0+$, implying that this shrunk surface does not represent a
black hole for $\varphi \neq $const. It is instead a naked singularity [34].
On the other hand, if $C\rightarrow 0$ and $\lambda \rightarrow 1$, we have
a finite value of $\mathbf{R}$ as $r\rightarrow B$. This means that we have
a black hole solution for $\varphi =$const, in total accordance with the
no-hair theorem.

Generally speaking, wormhole solutions obtain in the Jordan frame because
the sign of the energy density is indefinite in that frame. The sign is
positive or negative according as $C(\omega )+1<\lambda $ or $C(\omega
)+1>\lambda $. Let us examine the situation in the Einstein frame, defined
earlier.

\begin{center}
\textbf{IV. EINSTEIN FRAME}
\end{center}

Under the conformal transformation%
\begin{equation}
\widetilde{g}_{\mu \nu }=pg_{\mu \nu },\text{ \ \ }p=\frac{1}{16\pi }\varphi
,
\end{equation}%
and a redefinition of the Brans-Dicke scalar%
\begin{equation}
d\phi =\left( \frac{\omega +3/2}{\alpha }\right) ^{1/2}\frac{d\varphi }{%
\varphi },
\end{equation}%
in which we have intentionally introduced an arbitrary parameter $\alpha $,
the action (3) in the Einstein variables ($\widetilde{g}_{\mu \nu },\phi $)
becomes%
\begin{equation}
S=\int d^{4}x(-\widetilde{g})^{1/2}\left[ \widetilde{\mathbf{R}}-\alpha
\widetilde{g}^{\mu \nu }\phi _{,\mu }\phi _{,\nu }\right] .
\end{equation}%
The field equations are%
\begin{eqnarray}
\widetilde{\mathbf{R}}_{\mu \nu } &=&\alpha \phi _{,\mu }\phi _{,\nu } \\
\square ^{2}\phi &=&0.
\end{eqnarray}%
The solutions of Eqs. (30) and (31) can be obtained, using the
transformations (27) and (28), as%
\begin{eqnarray}
d\tau ^{2} &=&-\left( 1+\frac{B}{r}\right) ^{-2\beta }\left( 1-\frac{B}{r}%
\right) ^{2\beta }dt^{2}+\left( 1-\frac{B}{r}\right) ^{2(1-\beta )} \\
&&\times \left( 1+\frac{B}{r}\right) ^{2(1+\beta )}[dr^{2}+r^{2}(d\theta
^{2}+\sin ^{2}\theta d\psi ^{2})]  \nonumber
\end{eqnarray}

\begin{equation}
\phi =\left[ \left( \frac{\omega +3/2}{\alpha }\right) \left( \frac{C^{2}}{%
\lambda ^{2}}\right) \right] ^{1/2}\ln \left[ \frac{1-\frac{B}{r}}{1+\frac{B%
}{r}}\right] ,
\end{equation}%
\begin{equation}
\beta =\frac{1}{\lambda }\left( 1+\frac{C}{2}\right) .
\end{equation}%
The expression for $\lambda ^{2}$, of course, continues to be the same as
Eq. (9), and using this, we can rewrite Eq. (33) as%
\begin{equation}
\phi =\left[ \frac{2(1-\beta ^{2})}{\alpha }\right] ^{1/2}\ln \left[ \frac{1-%
\frac{B}{r}}{1+\frac{B}{r}}\right] .
\end{equation}%
Casting the metric (32) into the Morris-Thorne form, we can find the
wormhole throat $r$ radii to be%
\begin{equation}
r_{0}^{\pm }=B\left[ \beta \pm \sqrt{\beta ^{2}-1}\right] .
\end{equation}%
For real $r_{0}^{\pm }$, we must have $\beta ^{2}\geq 1$. But $\beta ^{2}=1$
corresponds to a nontraversable wormhole since $r_{0}^{\pm }$ coincides with
the singular radius $r_{S}=B$. From Eq. (35), it follows that, if $\alpha >0$
and $\beta ^{2}>1$, then no wormhole is possible as $\phi $ becomes
imaginary. This result is quite consistent with the fact that the
stress-energy tensor for massless minimally coupled scalar field $\phi $:
viz.,%
\begin{equation}
T_{\mu \nu }=\alpha \left[ \phi _{,\mu }\phi _{,\nu }-\frac{1}{2}\widetilde{g%
}^{\mu \nu }\varphi _{,\sigma }\varphi ^{,\sigma }\right]
\end{equation}%
satisfies all energy conditions [6]. The Einstein frame is thus called
\textquotedblleft physical\textquotedblright\ for which the restriction $%
\omega >-3/2$ follows from Eq. (33).

On the other hand, if we choose $\alpha <0$, which amounts to violating all
energy conditions by brute force, one may find wormholes for $\beta ^{2}>1$
in Eq. (35) or, equivalently, for $\omega <-3/2$.

We wish to point out a few more relevant points.

(i) Just as in the Jordan frame, the \textquotedblleft no-hair
theorem\textquotedblright\ holds also in the Einstein frame. This can be
seen from the expression for scalar curvature $\widetilde{\mathbf{R}}$
computed from the metric (32):%
\begin{equation}
\widetilde{\mathbf{R}}=\frac{8B^{2}r^{4}(1-\beta ^{2})}{(r-B)^{2(2-\beta
)}(r+B)^{2(2+\beta )}}.
\end{equation}%
One can see that $\widetilde{\mathbf{R}}$ is negative for wormhole
solutions. In the Schwarzschild limit $\beta \rightarrow 1$, $\widetilde{%
\mathbf{R}}$ is finite for $r\rightarrow B$, and a black hole solution
results, in complete accordance with the no-hair theorem [34]. The
divergence of $\phi $ at $r=B$ has been shown to be physically innocuous
[35,36]. Generally, for $\beta \neq 1$, $\widetilde{\mathbf{R}}\rightarrow
\infty $ as $r\rightarrow B$. This implies that the surface $r=B$ (or, $R=0$%
) is not a black hole surface for nonconstant $\phi $. This conclusion is in
agreement with that reached by Agnese and La Camera [37] in a different way.

(ii) The Arnowitt-Deser-Misner \symbol{126}ADM! mass of the configuration is
defined by%
\begin{equation}
M=\frac{1}{16\pi }\int_{S}\dsum\limits_{i,j=1}^{3}\left( \partial
_{j}g_{ij}-\partial _{i}g_{ii}\right) n^{i}dS,
\end{equation}%
where $S$ is a 2-surface enclosing the active region and $n^{i}$ denotes the
unit outward normal. For the metric (32), we get%
\begin{equation}
M=2B\beta ,
\end{equation}%
and using this value, the metric can be expanded in the weak field as%
\begin{eqnarray}
d\tau ^{2} &=&-(1+2Mr^{-1}+...)dt^{2} \\
&&+(1+2Mr^{-1}+2Mr^{-2}...)[dr^{2}+r^{2}(d\theta ^{2}+\sin ^{2}\theta d\psi
^{2})];  \nonumber
\end{eqnarray}%
that is, it predicts exactly the same results for a neutral test particle as
does Einstein's general relativity. The factor $\alpha $ does not appear in
the metric, although it does appear in the scalar field $\phi $. Hence, $%
\alpha $ cannot be determined by any metric test of gravity.

(iii) It should be remarked that if we replace $B$ by another integration
constant $m/2$, the solutions (32) and (35) become those proposed by
Buchdahl [38] long ago. Defining the field strength \ $\sigma $ for the
scalar field $\phi $ in analogy with an \textquotedblleft electrostatic
field,\textquotedblright\ one obtains%
\begin{equation}
\sigma =-2m\delta \text{, \ \ \ \ \ \ \ \ }\delta =[(1-\beta ^{2})/2\alpha
]^{1/2}.
\end{equation}%
Then, from Eqs. (40) and (42), it follows that the gravitation producing
mass $M$ is given by%
\begin{equation}
M^{2}=m^{2}-\frac{1}{2}\alpha \sigma ^{2},
\end{equation}%
where $m$ can be regarded as the strength of the source excluding the scalar
field. For $\beta \rightarrow 0$, we have $M\rightarrow 0$. The situation in
this case is that, for $\alpha >0$, we can have both $m$ and $\sigma $
nonzero, but with their effects mutually annulled. In other words, we obtain
a configuration which is indifferent to a gravitational interaction with
distant bodies. The reason is that the stresses of the field contribute an
amount of negative gravitational potential energy (attractive) just
sufficient to make the total energy zero [38]. On the other hand, if $\alpha
<0$, the $\phi $ field has a positive gravitational potential energy
(repulsive). We cannot take $\beta \rightarrow 0$ owing to Eq. (42), but it
is possible to make $m\rightarrow 0$ so that $M\rightarrow 0$. In this case,
we have $\sigma =0$. That is, the vanishing of total energy implies a
vanishing of individual source contributions.

\begin{center}
\textbf{V. SUMMARY}
\end{center}

The foregoing analysis reveals that spherically symmetric static vacuum
Brans-Dicke wormholes exist in the Jordan frame only in a very narrow
interval $-3/2<\omega <-4/3$, corresponding to a physical situation where
the post-Newtonian approximation is valid. In general, the wormhole range
for $\omega $ depends entirely on the form of $C(\omega )$ supposed to be
dictated by physical conditions. Wormhole solutions do not exist at all in
the conformally rescaled (Einstein) frame unless one is willing to violate
the energy conditions by choice ($\alpha <0$). However, such a manipulation
is not always necessary. For example, there exist theories where one adds to
the Einstein frame vacuum action other fields (such as the axion field [15])
or potentials [39] and obtains dynamic wormhole solutions in a natural way.

It is evident that the factor $\alpha $ does not appear in the
metric (32), although it does appear in the expression for the
scalar field $\phi $. In particular, for local tests of gravity,
the predictions are exactly the same as those of Einstein's
general relativity where the Robertson parameters take on values
$\alpha =\beta =\gamma =1$. In contrast, in the Jordan frame,
one has $\alpha =\beta =1$, $\gamma =(\omega +1)/(\omega +2)$. For finite $%
\omega $, it is evident that the predictions deviate somewhat from the
actually observed values.

The Arnowitt-Deser-Misner (ADM) mass of the configuration is positive in
both the frames. In the Jordan frame, it is $M=(2B/\lambda )(C+1)$, while in
the Einstein frame it is $M=2B\beta $. It is also shown that a
gravitationally indifferent real configuration with zero total energy ($M=0$%
) does or does not exist in the Einstein frame according as $\alpha >0$ or $%
\alpha <0$.

An interesting feature of Brans-Dicke wormholes is that infinitely large
radial tidal accelerations occur at the throat so that these wormholes are
not traversable in practice. This feature is reflected in the absence of a
black hole surface at $r=B$ or, in the Morris-Thorne coordinates, at $R=0$.

We have not addressed the question of stability of Brans- Dicke wormholes in
this paper. With regard to classical perturbations, it should be pointed out
that the results of Anchordoqui, Bergliaffa, and Torres [22] indicate that
addition of extra ordinary matter does not destroy the wormhole. The effect
of the quantum back reaction of the scalar field on stability will be
considered elsewhere.

\textbf{ACKNOWLEDGMENTS}

One of us (S.M.K.A.) would like to thank the Indian Council for Cultural
Relations (ICCR), New Delhi, for financial support under an Exchange Program
of the Government of India.

\textbf{REFERENCES}

[1] M. S. Morris and K. S. Thorne, Am. J. Phys. \textbf{56}, 395 (1988).

[2] M. S. Morris, K. S. Thorne, and U. Yurtsever, Phys. Rev. Lett. \textbf{61%
}, 1446 (1988).

[3] J. G. Cramer \textit{et al}., Phys. Rev. D \textbf{51}, 3117 (1995).

[4] D. Hochberg and T. W. Kephart, Gen. Relativ. Gravit. \textbf{26}, 219
(1994); Phys. Lett. B \textbf{268}, 377 (1991); Phys. Rev. Lett. \textbf{70}%
, 2665 (1993).

[5] S. Coleman, Nucl. Phys. \textbf{B310}, 643 (1988); \textbf{B307}, 867
(1988); S. W. Hawking, \textit{ibid}. \textbf{B335}, 155 (1990); S. Giddings
and A. Strominger, \textit{ibid}.\textbf{\ B321}, 481 (1988).

[6] M. Visser, Phys. Rev. D \textbf{39}, 3182 (1989); Nucl. Phys. \textbf{%
B328}, 203 (1989); \textit{Lorentzian Wormholes---From Einstein To Hawking }%
(AIP, New York, 1995).

[7] L. H. Ford and T. A. Roman, Phys. Rev. D \textbf{51}, 4277 (1995);
\textbf{53}, 5496 (1996); \textbf{53}, 1988 (1996); \textbf{55}, 2082 (1997).

[8] D. Hochberg, A. Popov, and S. V. Sushkov, Phys. Rev. Lett. \textbf{78},
2050 (1997); F. Schein and P. C. Aichelberg, \textit{ibid}. \textbf{77},
4130 (1996).

[9] W. G. Unruh, in \textit{Quantum Theory of Gravity}, edited by S. M.
Christensen (Hilger, Bristol, 1984).

[10] D. N. Page and C. D. Geilker, Phys. Rev. Lett. \textbf{47}, 979 (1981).

[11] T. Padmanabhan, T. R. Seshadri, and T. P. Singh, Int. J. Mod. Phys.
\textbf{A1}, 491 (1986).

[12] K. K. Nandi, A. Islam, and J. Evans, Int. J. Mod. Phys. \textbf{A12},
3171 (1997).

[13] L. Parker and S. A. Fulling, Phys. Rev. D \textbf{7}, 2357 (1973).

[14] D. Hochberg, Phys. Lett. B\textbf{\ 251}, 349 (1990).

[15] D. H. Coule and K. Maeda, Class. Quantum Grav. \textbf{7}, 955 (1990);
P. F. Gonzalez-Diaz, Phys. Lett. B \textbf{233}, 85 (1989); H. Fukutaka, K.
Ghoruku, and K. Tanaka,\textit{\ ibid}. \textbf{222}, 191 (1989).

[16] S. Cotsakis, P. Leach, and G. Flessas, Phys. Rev. D \textbf{49}, 6489
(1994).

[17] S. Giddings and A. Strominger, Nucl. Phys. \textbf{B306}, 890 (1988).

[18] A. Zee, Phys. Rev. Lett. \textbf{42,} 417 (1979).

[19] F. S. Accetta, A. Chodos, and B. Shao, Nucl. Phys. \textbf{B333}, 221
(1990).

[20] A. G. Agnese and M. La Camera, Phys. Rev. D \textbf{51}, 2011 (1995).

[21] K. K. Nandi, A. Islam, and J. Evans, Phys. Rev. D \textbf{55}, 2497
(1997).

[22] L. Anchordoqui, S. P. Bergliaffa, and D. F. Torres, Phys. Rev. D
\textbf{55}, 5226 (1997).

[23] N. Riazi and H. R. Askari, Mon. Not. R. Astron. Soc. \textbf{261}, 229
(1993).

[24] K. K. Nandi and A. Islam, Ind. J. Phys. B \textbf{68}, 539 (1994).

[25] T. Harada et al., Phys. Rev. D \textbf{55}, 2024 (1997).

[26] G. Magnano and L. M. Sokol\textquotedblright owski, Phys. Rev. D
\textbf{50}, 5039 (1994).

[27] C. H. Brans, Phys. Rev. \textbf{125}, 2194 (1962).

[28] T. Matsuda, Prog. Theor. Phys. \textbf{47}, 738 (1972).

[29] R. Ruffini and J. A. Wheeler, Phys. Today \textbf{24}(1), 30 (1971).

[30] S. W. Hawking, Commun. Math. Phys. \textbf{25}, 167 (1972).

[31] J. E. Chase, Commun. Math. Phys. \textbf{19}, 276 (1970).

[32] C. Teitelboim, Lett. Nuovo Cimento \textbf{3}, 326 (1972).

[33] J. D. Bekenstein, Phys. Rev. Lett. \textbf{28}, 452 (1972).

[34] A. Saa, J. Math. Phys. \textbf{37}, 2346 (1996).

[35] T. Zannias, J. Math. Phys. \textbf{36}, 6970 (1995).

[36] J. D. Bekenstein, Ann. Phys. (N.Y.) \textbf{91}, 72 (1975); \textbf{82}%
, 535 (1974).

[37] A. G. Agnese and M. La Camera, Phys. Rev. D \textbf{31}, 1280 (1985).

[38] H. A. Buchdahl, Phys. Rev. \textbf{115}, 1325 (1959).

[39] J. D. Bekenstein, Phys. Rev. D \textbf{51}, R6608 (1995).

\end{document}